# Properties of Building Blocks Comprising Strongly Interacting Posts and Their Consideration in Advanced Coaxial Filter Designs

Smain Amari, Mustafa Bakr and Uwe Rosenberg

*Abstract*—Building blocks containing strongly coupled posts offer new possibilities for advanced coaxial (comb-line) filter designs. Equivalent circuits based on the individual resonances of the posts cannot be used to reliably describe the behavior of these structures because of the strong coupling between the posts. Instead, sets of electromagnetic (EM) resonances that satisfy the boundary conditions are used. The resulting equivalent circuit is either a fully transversal circuit or contains locally transversal sub-circuits depending on the strength of the coupling between the cascaded blocks. The validity of similarity transformations that result in topologies with unusual strong coupling coefficients is questionable despite the fact that they yield the correct frequency response. Such coupling matrices obscure the physics of the problem and fail to predict the correct behavior of filtering structures. However, topologies that match the layout of the posts can be used to optimize the filter in connection with a *full-wave* solver or measurement. Examples of dual-post and triple-post units are used to illustrate the key findings. The basic knowledge of the real functionality of these special resonator configurations allows their consideration in advanced filter implementations by well-established classic design methods, without limitation by the design approach. This is demonstrated by an example of a 2-order in-line filter implementation providing one transmission zero by using the combination of single and transverse dual-post resonators. This fundamental understanding of the special properties provides the pre-requisite for a variety of novel filter solutions.

*Index Terms*—Coaxial filter, combline filter, coaxial resonator coupling transformation, zero shifting property, singlet, doublet, filter synthesis.

## I. INTRODUCTION

ADVANCED filter design aims at the realization of tailored filter functions to satisfy rejection demands with the lowest possible filter order, while accommodating low passband insertion loss and reasonable implementation expenses. Filtering functions with transmission zeros (TZ's) whose frequency locations are chosen to meet rejection requirements are exploited. Cross (bypass) coupling is the most commonly used mechanism to implement TZs in microwave filters, e.g. [1].[1] Placing a TZ at a given frequency with respect to the passband often imposes specific relations between the coupling coefficients, especially their relative signs. Building blocks are characterized by the maximum number of TZs they can generate and the relative signs of the required coupling coefficients for the generation of the TZ's at the corresponding locations.

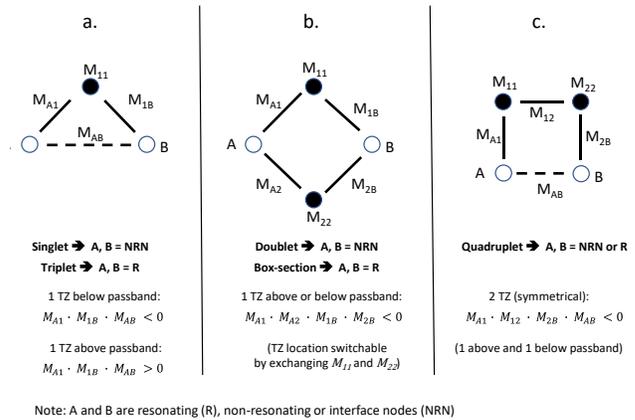

Fig. 1. Basic building blocks and pre-requisites (coupling signs) for the realization of TZs.

The majority of filter designs consider (combinations of) basic building blocks for the implementation of TZs, like doublets, triplets and quadruplets, since they allow the individual control of dedicated TZs in an overall (higher order) filter design (cf., Fig. 1). For the implementation of TZs, they rely on combinations of couplings with inductive ('positive') and capacitive ('negative') nature.

Classical coaxial filter designs are based on sequentially arranged $\lambda/4$ resonators[2] (posts) within a housing. The coupling of adjacent resonators is generally of inductive nature due to the dominant magnetic field components at the bottom of the housing while there are also electric field components at the top (open end) of the resonators. Thus, a partial wall between adjacent resonators may be used to control the dedicated couplings; i.e., a partial wall at the top yields an increase of the inductive coupling; whereas a partial wall at the bottom suppresses the dominant inductive coupling and thus will yield a weak capacitive coupling between the adjacent resonators. For stronger capacitive couplings special probe designs are usually considered that are supported in the separation wall between the adjacent resonators (e.g., [2,3]). Obviously, the implementation of such coupling designs is more elaborate and sensitive especially in the case of strong couplings yielding small distances between resonators and probes. This may also degrade the power handling capability of the filter. Consequently, solutions that avoid any probe couplings for the implementation of needed negative couplings are preferred. This can be obtained principally by the utilization of transformation properties, such as phase reversal, of certain resonator implementations as introduced for waveguide cavity filters in [4].

---

[1] TZs may be also realized with extracted poles. Since such special implementations are not relevant regarding the introduction of the basic properties of close post arrangements, they are not further mentioned.

[2] The resonators may be shorter than $\lambda/4$ depending on capacitive loading at the top, e.g., gap towards the top wall.

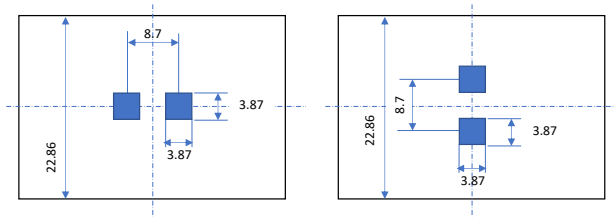

Fig. 2. Dimensions of dual-post configurations for analyses of basic behavior; left: inline; right: transverse (housing height 22.86mm, height of posts 22.00mm)

A building block, which contains two strongly coupled identical posts was introduced in [5,6]. Only the odd mode of the block was used in the design of the passband; the even mode resonates far away from (below) the passband and was considered spurious. The odd-mode nature of this configuration was utilized for the implementation of TZs below and above the passband. The same idea was used in a triple-post building block [7-9]. This configuration exhibits two close resonances, which are used in the passband, and the third resonance that is located far away (below) the passband is considered spurious. TZ's below and above the passband can be implemented as the results in [7-9] show. While closely placed posts for filter applications were first reported more almost three decades ago by Rosenberg et.al. [10], they have been used in different and interesting ways in the recent literature [5-9].

A common feature of the dual-post and the triple-post building blocks analyzed in [5-9] is the presence of strong coupling between at least two posts. Consequently, the number of passband resonances provided by each building block is one less than the number of posts. This observation suggests that associating a resonance to each post, as is the case in some coupling schemes, should be viewed with suspicion.

A transversal network, which is also known as the admittance representation in the theory of linear systems, and its narrow-band limit[3] that is commonly known as the transversal coupling matrix, use the eigen-resonances of the complete structure with all the perturbations present as a basis. Although it provides a general and universal representation of coupled resonator filters, the transversal coupling matrix is not convenient for initial designs. Instead, similarity transformations (rotations) which preserve the symmetry of the coupling matrix and its frequency response are then used to generate new coupling schemes with a topology that matches the layout of the individual physical and localized 'resonators'. Such transformations, which are perfectly valid at the coupling matrix level, are questionable for real structures with strong couplings between resonators whose EM fields share the same volume mainly because of the violation of the boundary conditions. It is important not to lose sight of the fact that the actual filter is described by vector quantities, the components of the electromagnetic field, that obey a set of coupled partial differential equations, Maxwell's equation, with associated boundary conditions. The coupling matrix represents a system that obeys at set of coupled first order ordinary differential equations with no knowledge of the boundary conditions. The concept of equivalence between the two must be viewed with care and suspicion, especially when strong coupling coefficients are present.

The following section of the paper discusses the concept of similarity transformations and the implication of boundary conditions. This is followed by an analysis of dual-post and triple-post resonators in order to understand their properties. The use of these building blocks in filter designs is then addressed. The understanding of the characteristics of the close-post resonators allows the application of classical well-established filter design methods without limitations, even when combining different types of resonators. This is demonstrated by way of an example of a 2-order in-line filter design providing one TZ by using the combination of single and transverse dual-post resonators. The dimensions obtained by a pre-design with a classical design approach yields already accurate performance results. In addition, examples of novel building blocks and 3$^{rd}$ and 4$^{th}$ order filters are also introduced to confirm the results of the standard design methods.

## II. BOUNDARY CONDITIONS AND SIMILARITY TRANSFORMATIONS

We consider the structures shown in Fig. 2. They consist of two closely placed posts inside a metallic enclosure. We assume that the structure is lossless and homogenous. The posts are identical with a gap between their top and the metallic cover. The eigen-resonances of the structures can be straightforwardly determined by using a modern full-wave solver (e.g. [11]). The field distributions of the two lowest eigen-resonances of the transverse dual-post structure (Fig. 2, right) are shown in Figs. 3 and 4[4]. The EM fields of the resonances are not concentrated around any of the two posts.

These two resonances are not coupled to each other as required by the orthogonality of the solutions of Maxwell's equations in a given volume. Because of the symmetry of the structure, we denote the even mode by $\emptyset_e(\vec{r})$ and the odd mode by $\emptyset_o(\vec{r})$. The quantity $\emptyset_{e,o}(\vec{r})$ can be taken as a component of the electric field, magnetic field or a relevant modal electromagnetic quantity depending on the coordinate system. These two functions are eigen-functions of a linear operator $\mathcal{L}$ along with the associated boundary conditions. If the eigenvalues are $\lambda_e = \omega_e^2$ and $\lambda_o = \omega_o^2$, we have

$$\mathcal{L}\emptyset_e = \omega_e^2 \emptyset_e \quad (1.a)$$

$$\mathcal{L}\emptyset_o = \omega_o^2 \emptyset_o \quad (1.b)$$

---

[3] The meaning of 'narrow band' is related to the validity in view of the dedicated physical implementation and passband; note, there are also special broadband resonator filter designs relying on this representation, but it is only valid for the realizable characteristics close to the passband.

[4] EM calculations and design analyses conducted with µWaveWizard™ from Mician (https://www.mician.com)

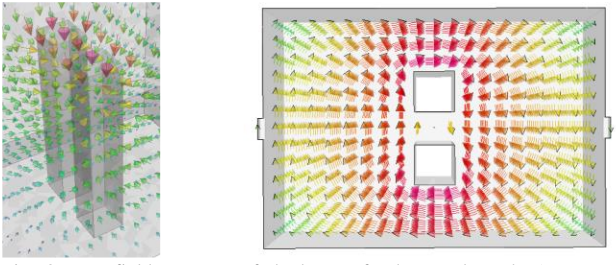

Fig. 3. EM field patterns of dual post fundamental mode (resonance frequency: 2.095GHz), left – zoom of E-fields, right – H-fields, cut at half housing height.

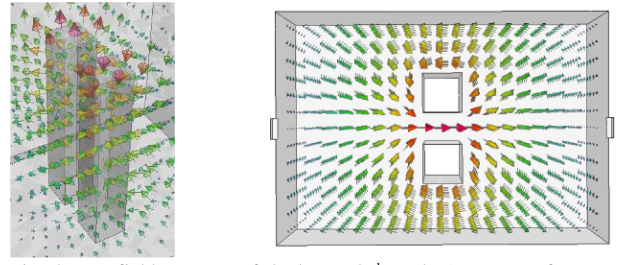

Fig. 4. EM field patterns of dual post 2nd mode (resonance frequency: 2.781GHz), left – zoom of E-fields, right – H-fields, cut at half housing height.

If the input and output ports that couple to both resonances are added to the structure, we get the equivalent circuit shown in Fig. 1b when the resonances $\emptyset_e(\vec{r})$ and $\emptyset_o(\vec{r})$ are used as basis. It is a doublet.

Let us now consider the question of representing the circuit by using localized 'resonances' that are associated to each of the posts individually. We define the following two functions

$$\emptyset_1 = \frac{1}{\sqrt{2}}(\emptyset_e + \emptyset_o) \quad (2.a)$$

$$\emptyset_2 = \frac{1}{\sqrt{2}}(\emptyset_e - \emptyset_o) \quad (2.b)$$

When the structure is represented in the new basis $(\emptyset_1, \emptyset_2)$, we get the coupling scheme shown in Fig. 5.

The transformation described by equations (2.a) and (2.b) is a similarity transformation, in this case a rotation by 45 degrees.

The first important point that we need to address is whether the quantities $\emptyset_1$ and $\emptyset_2$ are eigen-resonances. For this to be the case, they must satisfy the eigen-problem in equations (1). We have

$$\mathcal{L}\emptyset_1 = \frac{1}{\sqrt{2}}(\mathcal{L}\emptyset_e + \mathcal{L}\emptyset_o) = \frac{1}{\sqrt{2}}(\omega_e^2 \emptyset_e + \omega_o^2 \emptyset_o) \quad (3.a)$$

$$\mathcal{L}\emptyset_2 = \frac{1}{\sqrt{2}}(\mathcal{L}\emptyset_e - \mathcal{L}\emptyset_o) = \frac{1}{\sqrt{2}}(\omega_e^2 \emptyset_e - \omega_o^2 \emptyset_o) \quad (3.b)$$

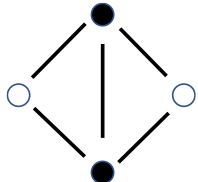

Fig. 5. Coupling scheme of dual-post with $\emptyset_1$ and $\emptyset_2$ in equations (2) used as basis

To get more insight into this transformation, we introduce the sum and difference

$$S = \omega_e^2 + \omega_o^2 \quad (4.a)$$

$$D = \omega_e^2 - \omega_o^2 \quad (4.b)$$

Combining equations (3) and (4), we get

$$\mathcal{L}\emptyset_1 = \frac{S}{2}\left(\emptyset_1 + \frac{D}{S}\emptyset_2\right) \quad (5.a)$$

$$\mathcal{L}\emptyset_2 = \frac{S}{2}\left(\emptyset_2 + \frac{D}{S}\emptyset_1\right) \quad (5.b)$$

It is evident that these two functions are not eigen-functions unless $D/S = 0$, or equivalently that the two original modes are degenerate, i.e., $\omega_e = \omega_o$. However, if $D/S$, which is nothing other than the coupling coefficient between the two posts, is small enough to be neglected, the two functions are approximately eigen-functions and the eigen-functions are approximately degenerate. This is the case of narrow-band coupled resonator filters. When $D/S$ is large, as in strongly coupled resonators, the error is treating $\emptyset_1$ and $\emptyset_2$ as eigen-functions (resonances) may be large enough to render a coupling matrix based on these non-physical local resonances unreliable, especially when these share the same volume because of the orthogonality properties of solutions of Maxwell's equations. The elements of the resulting coupling matrix will not be independent because of the boundary conditions. Paths going through such 'resonances' are not allowed to transport energy independently. Predictions based on coupling schemes which use these localized 'resonances' are not always reliable. For example, a triplet of three identical cylindrical TEM posts, with predominantly magnetic coupling, have local 'resonances' with magnetic fields that result in a negative coupling coefficient thereby predicting a TZ below the band. This is not what is observed experimentally or in full-wave simulation.

An interesting by-product of this analysis is the appearance of a coupling coefficient. Indeed, equations (5) show that the relative amplitudes of the functions $\emptyset_1$ and $\emptyset_2$ are related by

$$k = \frac{D}{S} = \frac{\omega_e^2 - \omega_o^2}{\omega_e^2 + \omega_o^2} \quad (6)$$

An identical expression of the coupling coefficient between two resonators was given by Awai in a recent article [12]. The advantage of the derivation in this article is its generality since it applies to any two-state system that is linear and invariant under time reversal so that the eigenvalues are the squares of the angular frequencies.

When the 'resonators' share the same volume, it is important to use those solutions that satisfy the boundary conditions in predicting the behavior of the filter, especially when strong inter-resonator coupling is present as in the case of strongly coupled dual and triple posts. These are the only separate paths that the energy can follow between the input and the output and contain all the relevant physics of the problem assuming that all other resonances are far enough from the frequency range of interest. Certain properties of the filter that are determined by the signals following these different paths are obscured by similarity transformations. For example, it is known that the TZ of a doublet, or a box-section, can be shifted from one side of the passband to the other by changing the signs of the diagonal elements of the coupling matrix [13]. Although similarity transformations preserve the response of the filter, they do not preserve the TZ-shifting property. Instead, similarity transformations result in coupling schemes that obscure this important feature that is exploited in moving the TZ from one side of the passband to the other.

At this point one might wonder why the coupling matrix whose topology matches the layout of the localized 'resonators' is used successfully in designing these filters even when strong inter-resonator coupling is present [7-9]. Here, it is important to understand that this is the case only for optimization-based design methods that rely on the frequency response of the filter (or dedicated subsection) that is obtained from a *full-wave* solver. The full-wave simulator *guarantees* that the boundary conditions are all satisfied. As long as the structure is capable of implementing a response in the range of the coupling matrix whose elements are assumed reliably extracted from the full-wave solution, the process should be successful. Yielding the desired response does not mean that the coupling matrix represents the local physics of the problem accurately. All matrices that are related by similarity transformations yield the same response but this does not mean they represent the local physics of the problem equally well. The coupling matrix that is based on localized resonators is 'helped' by the fact that not all coupling coefficients are strong, meaning that parts of filters are adequately represented by localized resonances. However, this coupling matrix cannot represent correctly the physics of individual sections which contain strong couplings such as dual-post and triple-post units. Consequently, the classic design technique that is based on isolated pairs of localized resonators, and which provides a more rigorous test of the localized resonator concept, will fail in case of strong coupling coefficients between 'resonators'. This failure, which is familiar to filter designers, was also reported in [6].

The important conclusions of this analysis are the following:
a) The physics of the structure is accurately represented by those resonances that satisfy all the boundary conditions.
b) Similarity transformations do not define new valid local 'resonances' because of the violation of the boundary conditions, when they result in strong inter-resonator coupling. Coupling matrices with such strong coupling coefficients do not describe correctly the characteristics of those sections of the filter that contain resonators that share the same volume although they produce the correct overall frequency response (by construction).
c) Similarity transformations define approximate resonances which are valid representations for filters with narrow bandwidths or weak inter-resonator coupling where the resonances, which satisfy the boundary conditions are approximately degenerate.
d) The frequency response of the filter as obtained from a *full-wave* solver or measurement can be used in connection with a coupling matrix whose topology matches the layout of the localized 'resonators' to optimize the filter even when strong coupling coefficients are present. The classic design approach that is based on pairs of localized 'resonators' will fail when strong inter-resonator coupling coefficients are present [6].

We next apply these ideas to the cases of strongly coupled dual- and triple-posts.

III. PROPERTIES OF DUAL-POST UNIT

*A. Transversal dual-post unit*

A top view of the analyzed structure is shown in Fig. 2, right. We assume that the coupling between the unit and the input and output is weak enough for the modal field distributions not to be significantly affected. This is the case for narrow-band filters. With this assumption, the coupling scheme is a classic doublet (Fig. 1b) where the even and odd resonances of the dual-post units are taken as resonances 1 and 2. We use the odd resonance in the passband and call the even out-of-band resonance, resonance 2, spurious. The other choice is naturally possible.

A straightforward analysis of the coupling scheme in Fig. 1b shows that a TZ appears at the normalized frequency

$$\omega_z = -\frac{M_{11}+pM_{22}}{1+p}, \quad p = \frac{M_{S1}M_{1L}}{M_{S2}M_{2L}} \quad (7)$$

If we introduce the normalized resonance frequencies of the even and odd modes, $\omega_{od} = -M_{11}$ and $\omega_{sp} = \omega_e = -M_{22}$, we can re-write equation (7) in the form

$$\omega_z = \frac{\omega_{od}+p\omega_{sp}}{1+p}, \quad p = \frac{M_{S1}M_{1L}}{M_{S2}M_{2L}} \quad (8)$$

From this equation, we obtain the distance from the TZ to the two resonances as

$$\omega_z - \omega_{sp} = \frac{1}{1+p}(\omega_{od} - \omega_{sp}) \quad (9.a)$$

$$\omega_z - \omega_{od} = -\frac{p}{1+p}(\omega_{od} - \omega_{sp}) \quad (9.b)$$

$$\frac{|\omega_z-\omega_{od}|}{|\omega_z-\omega_{sp}|} = |p| \quad (9.c)$$

From these equations, assuming that the odd mode resonates above the even mode, i.e., $\omega_{od} - \omega_{sp} > 0$, we establish the following results

1. $0 < p < 1$. The TZ is located between the two resonances and is closer to the odd resonance than the spurious even resonance.
2. $p = 1$. The TZ is located at the average of the two resonances.
3. $p > 1$. The TZ is located between the two resonances and is closer to the spurious even resonance

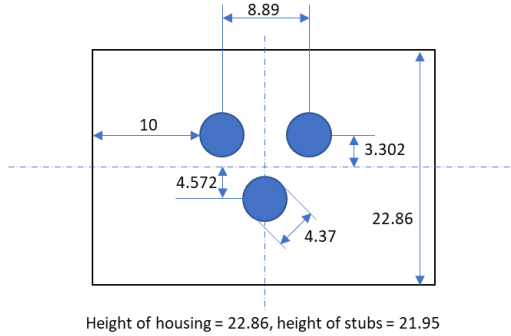

Fig. 6. Dimensions of triangular post configurations for eigenmode analyses (footprint and housing height according to [7], Fig. 12).

4. $0 > p > -1$. The TZ is located above the two resonances.
5. $p = -1$. The TZ is located at infinity. This is an all-pole response.
6. $p < -1$. The TZ is located below the two resonances.

From the field distributions of the two modes of the dual-post unit (Figs. 3 and 4, respectively), we see that the coupling to the odd mode is weaker than the coupling to the even mode. In fact, the odd mode is not excited at all when the input and output ports are placed at the horizontal symmetry plane between the two posts (cf. Fig. 2, right). This gives the freedom needed to adjust the parameter $p$ in equation (7). More specifically, we have two cases:
1. Input and output ports on the same side of the symmetry plane of the dual-post unit. The coupling coefficients are such that $0 < p < 1$. The TZ is between the odd and the even resonances but is closer to the odd resonance, i.e., below the passband. Its distance to the passband is reduced by moving the input and output ports closer to the symmetry plane of the dual-post unit.
2. Input and output ports on opposite sides of the symmetry plane of the dual-post unit. The coupling coefficients are such that $-1 < p < 0$. The TZ is located above the passband. Its distance to the passband is reduced by moving the input and output ports closer to the symmetry plane of the dual-post unit.

In both cases, the TZ can be placed arbitrarily close to the passband regardless of the value of the resonance frequency of the spurious even mode.

*B. In-line dual-post unit*

The coupling scheme of the dual post configuration with a 90° alignment (cf. Fig. 2, left) is still a doublet. In this case the input and output form an in-line configuration with the two posts of the unit. Obviously, this EM-field distribution of the resonances are similar to that of Figs. 3 and 4 but rotated by 90°. Consequently, the field distributions of the two resonances at the outer sides of the two posts along a line joining the two posts are very similar except for a phase reversal for the odd mode. This means that the coupling coefficients to the two modes, mainly through the evanescent $TE_{10}$ mode in the uniform waveguide section, are approximately equal but with one of them negative. This corresponds to $p = -1$. As a result, this configuration generates a TZ that is far away from the passband or even at infinity. Note that a TZ might be generated if the higher order modes carry enough energy around the resonant frequency of the odd mode.

IV PROPERTIES OF TRIPLE-POST UNIT

We now consider a unit formed by a pair of strongly coupled posts and an additional post that is not necessarily as strongly coupled to the pair. The structure generates three resonances, two close to each other and 'contribute' two resonances to the passband, the third one is far away from the passband and called spurious. The pair of posts is placed in the longitudinal direction as shown in Fig. 6.

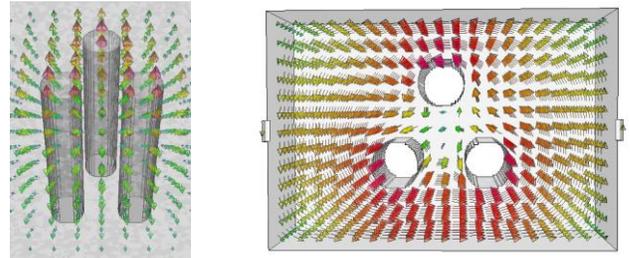

Fig. 7a. EM field patterns of fundamental mode - all three posts (resonance frequency: 2.057GHz), left – zoom of E-fields, right – H-fields, cut at half housing height.

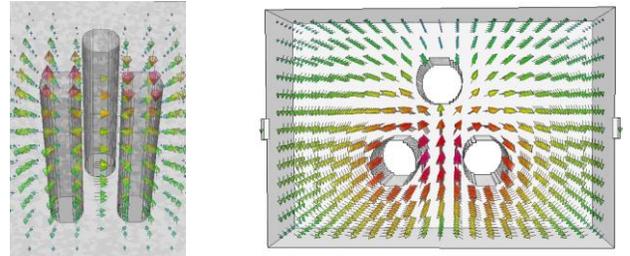

Fig. 7b. EM field patterns of 2$^{nd}$ (odd) resonance mode dedicated to both inline stubs (resonance frequency: 2.663GHz), left – zoom of E-fields, right – H-fields, cut at half housing height.

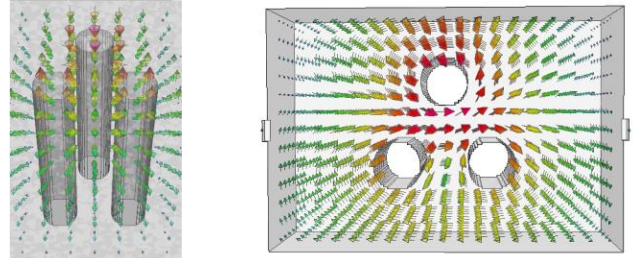

Fig. 7c. EM field patterns of 3$^{rd}$ (even) resonance mode dedicated to all three stubs (resonance frequency: 2.706GHz), left – zoom of E-fields, right – H-fields, cut at half housing height.

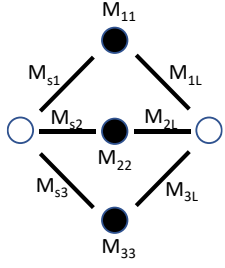

Fig. 8. Transversal coupling scheme of triple-post configuration.

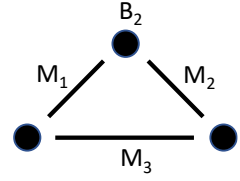

Fig. 9. Triple-post represented by localized 'resonances'.

First the EM fields distributions of the three resonances are determined; these are shown in Figs. 7 a, b and c. These resonances are *orthogonal* to one another (not coupled). It is obvious that none of these resonances is localized around any of the posts although the EM fields of the third one is approximately concentrated near the individual post. Consequently, a good model of the structure is the transversal coupling scheme shown in Fig. 8. We assume that spurious resonance, taken as resonance 3 (Fig. 7a), is far below the passband.

From the field distributions, we see that there are two even modes and one odd mode. The symmetry is with respect to the axis going through the individual post and the symmetry plane of the dual-post unit. This means that one coupling coefficient to the odd mode (either to the input or the output) can be assumed negative with the remaining ones all positive.

This unit generates up to two TZ's at finite frequencies. To generate only one TZ, we must impose constraints on its coupling coefficients. The generated TZ is located at

$$\omega_z = \frac{\omega_1 + p\omega_2}{1+p} + \mathcal{O}\left(\frac{1}{\omega_{sp}}\right), |\omega_{sp}| \gg 1 \quad (10.a)$$

with

$$p = \frac{M_{s1}M_{1L}}{M_{s2}M_{2L}}, \omega_1 = -M_{11}, \omega_2 = -M_{22}, \omega_{sp} = -M_{33} \quad (10.b)$$

and

$$M_{s1}M_{1L} + M_{s2}M_{2L} + M_{s3}M_{3L} = 0 \quad (10.c)$$

Equation (10.c) is the constraint needed to generate only one TZ at a finite frequency. The notation in equation (10.a) is used to indicate that the effect of the spurious resonance on the location of the TZ decreases at least as the reciprocal of $\omega_{sp}$. The important point to note from this equation is that the position of the TZ is mainly determined by the two resonances that contribute to the passband. In other words, the position of the TZ is not very sensitive to the location of the spurious resonances, as long as it is not too close to the passband. This is contrary to the transverse dual-post unit where the location of the TZ is directly related to the frequency of the spurious resonance as shown by equation (8).

The parameter $p$ in equation (10.b) is negative because one of the coupling coefficients is negative. This means that the TZ is located either below the passband or above depending on the relative strengths of the coupling to the two resonances.

From equation (10.a) we also see that the position of the TZ can be shifted from one side of the passband to the other by simply changing the signs of $\omega_1$ and $\omega_2$. This TZ-shifting property is not exact since it affects the shape of the response but becomes more accurate as the spurious resonance moves away from the passband. The position of the TZ is indeed moved to the other side of the passband under this transformation as shown on actual designs in this paper.

V. ROLE OF THE SPURIOUS RESONANCE

The even (spurious) mode of the transverse dual-post unit is in principle necessary for the generation of the TZ. For the triple-post unit, the dominant even (spurious) mode is not necessary for the design of the filter. It is the price one pays for generating the odd mode with the required phase reversal by using this structure. It is best to place it as far away from the passband as possible, while considering other resonator design aspects like unloaded Q. Although its effect on the passband may not be significant, it is important to have a way of taking it into account.

One approach would be to use a model that is based on the non-physical resonances localized around each post by viewing the filter as a dual-band filter [7-9]. The triple-post unit is then represented by a triplet with unrealistic coupling coefficients as shown in Fig. 2 in [9]. Although the corresponding coupling matrix can be used to optimize the filter from a full-wave analysis, it obscures the physics of the problem. It does not represent correctly the behavior of the sections with strong coupling.

The elements of the coupling matrix corresponding to the topology in Fig. 9 are no longer independent because of the boundary conditions. Several elements are affected by the same quantity. For example, the location of a TZ generated by the triple-post unit is given by [9]

$$\omega_z = \frac{M_1 M_2}{M_3} - B_2 \quad (11)$$

An examination of the results given in Fig. 2 in [9] show that $B_2$ changes from 2.1214 to 1.2912 when the normalized location of the spurious resonance changes from -11.69 to -3.09. At the same time, Fig.4 in the same reference shows that the location of the TZ has barely moved (Note that $\omega_{sp}$ in equations (10) is not exactly equal to the spurious resonance of the 4th order filter in this example, but the difference between the two is not sizable). An examination of equation (11) shows that this is possible only if the coupling coefficients and $B_2$ all depend on the spurious resonance. Unfortunately, the nature of this dependence is completely lost after the series of similarity transformations. On the other hand, equation (10.a) shows clearly that the location of the TZ is indeed not sensitive to the frequency of the spurious resonance when it is sufficiently far from the passband.

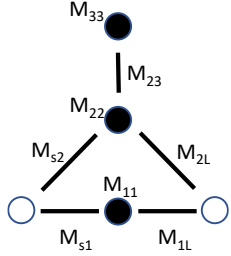

Fig. 10. Alternative coupling scheme of triple-post unit (after [9]).

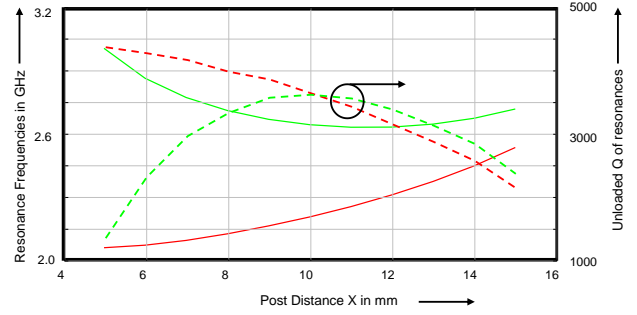

Fig. 11. Transverse dual-post properties according Fig.2 (right) for different post distance; red curves: dominant (even) mode; green curves: odd mode.

Another proposed coupling scheme of the triple-post unit is shown in Fig.10 (also shown as part of Fig. 7 in [9]). A straightforward analysis of this circuit shows that when only one transmission zero is generated, we have

$$\omega_z = -M_{33} + \frac{M_{23}^2}{M_{22}-M_{11}} \quad (12.a)$$

$$M_{s1}M_{1L} + M_{s2}M_{2L} = 0 \quad (12.b)$$

Here, resonator 1 is the odd resonance of the dual-post unit, resonator 2, the even mode and resonator 3, the single post. This model predicts that the TZ is completely determined by the single post ($M_{33}$) when the spurious resonance, $-M_{22}$, is far away from the passband. In the same limit, the structure produces only one return loss pole, instead of the desired two, because of a zero-pole cancellation. Naturally this result cannot be correct since the location of the TZ depends on the in-band eigen-resonances as given by equation (10.a) in this limit. This is another manifestation of using 'resonances' that do not satisfy the boundary conditions despite the fact the corresponding coupling schemes are admittedly very intuitive. Energy cannot flow through anything other than solutions to Maxwell's equations that satisfy the boundary conditions.

From a practical point of view, arguably the strongest reason for not including the spurious resonance in the model, especially for design purposes, is the impossibility of controlling it. Unfortunately, this is not the case. If we were able to control it, we would simply not couple to it and eliminate it. The spurious resonance plays a marginal role in the passband, only a small role in determining the location of the TZ of a triple-post unit. It should not be treated on par with the resonances that directly determine the passband performance of the filter. In this paper, we adjust the elements of the model that is based on the physical resonances that contribute to the passband to compensate for the effect of the spurious resonances on the passband and its vicinity. This approach has been used successfully for decades in waveguide technology.

## VI. CONSIDERATIONS FOR SYSTEMATIC FILTER DESIGN

The understanding of the real filter functionality allows basic design consideration for a systematic filter design like pre-investigations of the resonators in view of unloaded Q and spurious mode distance as well as predesign of couplings with assessment of initial structural dimensions. Moreover, it is a prerequisite for the consideration of well-established filter design methods providing systematic tuning of filter resonances and couplings in case of an EM CAD supported design (e.g., considering 'port tuning') and of course for the tuning of the final hardware. It is a basis for the application of well-established filter design methods without any general limitations.[5] The transverse dual-post resonator is used in the following as an example to introduce the basic filter design considerations when using resonators of close post configurations.

### A. Basic dual-post resonator design.

One essential aspect in filter design is the investigation of appropriate resonators in view of unloaded Q and spurious free bands. Thus, the respective properties of a transverse dual-post resonator according to Fig. 2 (right) have been explored for different post distances. The results in Fig. 11 show that the frequency distance between the dominant mode and the 2nd

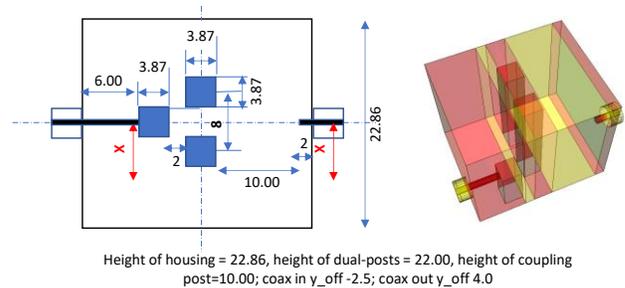

Height of housing = 22.86, height of dual-posts = 22.00, height of coupling post=10.00; coax in y_off -2.5; coax out y_off 4.0

Fig. 12. Configuration for input coupling investigation

---

[5] The approach in [6] does produce a final design that meets the specifications. However, it does not, provide a good initial design where single resonators and pairs of interacting resonators are designed as in the classic approach [6]. As mentioned in [6], it relies on an EM CAD 'port-tuning' method considering the subsection of the dual-post together with the adjacent single post resonators. This structure is optimized (tuned) to satisfy the characteristic between the ports, which of course corresponds to the original design, however, with an overdetermined equivalent circuit model, i.e., with a higher number of elements (5 couplings, 4 resonators) than that of the original design (3 couplings, 3 resonators). An optimization/'tuning' of the configuration with the higher number of variable physical parameters will finally satisfy the defined port characteristics. Moreover, the combination of the dual-post with the adjacent single post resonators yields a general design limitation, e.g., dual-post resonators may be also directly coupled without any intermediate single post resonator, also situated close to an interface coupling.

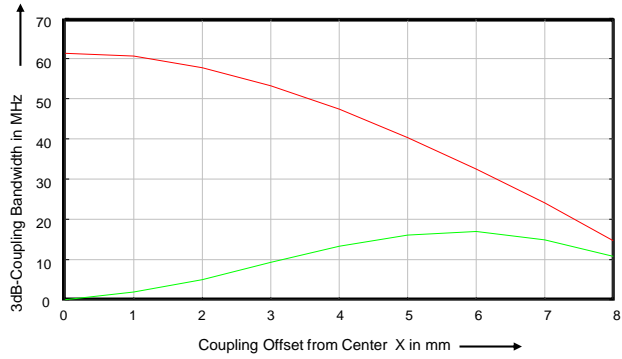

Fig. 13. Variation 3dB-coupling bandwidth of dual-post modes according to configuration in Fig. xx; red: basic mode (at 2.23GHz); green: 2nd mode (at 2.77GHz).

mode decreases with increasing post distance. In addition, the unloaded Q of the basic mode has a maximum for the close post distance, while it decreases with larger post spacing. In case of the 2nd mode the unloaded Q reaches a maximum at a some post spacing (center), approximately 10mm in this case. This is due to the stored EM energy between the posts for the 2nd mode causing higher surface currents for close distances. The decrease for larger spacings can be attributed to the increasing currents close to the side-wall, which holds also for the basic mode. For the application of the 2nd mode in a filter design, the post height and spacing can be used to achieve a compromise between the unloaded Q and the distance of the basic (spurious) mode. (The post footprint may also be considered for a final 'fine tuning' of these parameters).

### B. Consideration of resonator couplings

Another important filter design aspect is the suitable implementation of the resonator couplings. This holds for the in-/output couplings as well as for the inter-resonator ones.

*In-/output coupling design*

For the in-/output coupling investigation, a type of loop coupling is applied; i.e., a stub is located in front of the dual-post resonator, which is associated with the penetrating inner conductor of the coaxial interface at the end-wall (cf. Fig. 12)[6]. A weak probe coupling is considered with a convenient distance at the opposite end-wall for the indication of the 3dB coupling bandwidth, that is directly related to the dedicated input coupling and external Q, respectively. Evidently, an increase of the longitudinal distance between the coupling configuration and the resonator yields a decrease of the 3dB coupling bandwidth and thus the related effect on the dedicated coupling, In case of the dual-post resonator, the bandwidth variation has been explored when changing the coupling location in the transverse direction.

The coupling coefficients to each of the two resonances of the dual-post unit are shown in Fig. 13. In detail, the coupling to the dominant mode has a maximum (3dB-bandwidth) at the waveguide center; an increasing transverse center offset of the coupling location yields a decrease of the 3dB-bandwidth due to diminishing magnetic field components. In the case of the 2nd (odd) mode the 3dB bandwidth is obviously zero (no coupling) at the center location. However, an offset location yields a coupling of this mode with a maximum at a center offset of approximately 5.7mm for the used dimensions. These results are easily understood by noting that the coupling to the two resonances is determined by the evanescent $TE_{10}$ and $TE_{20}$ waveguide modes, respectively, in the waveguide section between the input stub and the dual-post unit. Obviously, the $TE_{10}$ mode with its lower cut-off frequency will always exhibit stronger coupling. The magnetic $H_x$ fields of the evanescent $TE_{20}$ mode vanishes at the center location while they have their maximum at the $1/4$ waveguide widths, which directly corresponds to the deduced couplings of the 2nd dual post mode (cf., Fig. 13).

---

[6] Other coupling configuration may be treated also in a similar manner.

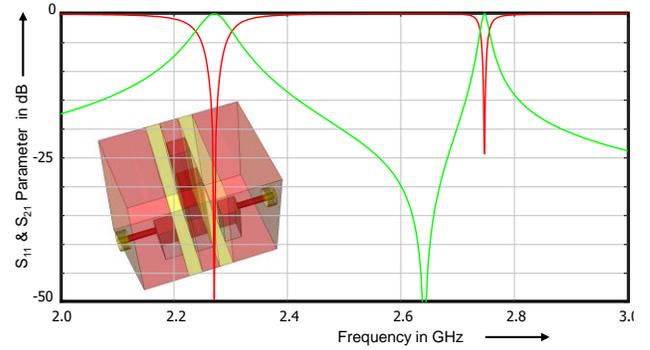

Fig. 14. Dual-post singlet; analyzed response with coupling stub offset in same direction (cf. inset)

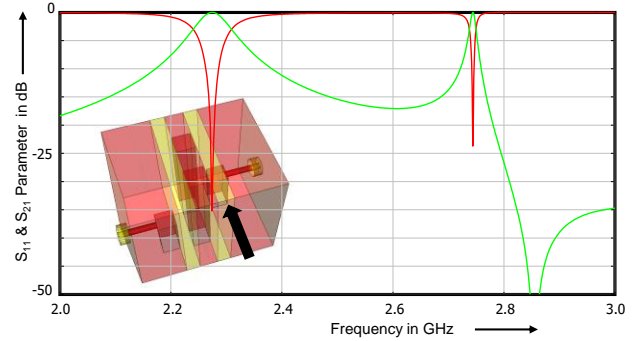

Fig. 15. Dual-post singlet; analyzed response with coupling stub offset in opposite direction (cf. inset)

*Coupling transformation properties / singlet*

The coupling transformation properties of the 2nd (odd) mode of the transverse dual-post unit, and its singlet-like behavior, can be easily verified with the configuration in Fig. 12, but using the couplings at input and output. The bypass coupling is mostly through the $TE_{10}$ evanescent mode of the waveguide that does not interact with the resonance (2nd mode of the dual post resonator) but it causes a bypass coupling between the input and output, while the basic spurious mode has negligible impact on the bypass coupling due to the large frequency distance.

When the input and output ports are placed on the same side of the symmetry plane of the configuration, as shown in Fig. 14 (inset), the generated TZ is below the resonance frequency of the odd mode. The full-wave results shown in Fig. 14 confirm this statement.

To move the TZ above the resonant frequency of the odd mode, the input and output ports must be placed on opposite sides of the symmetry plane of the posts as shown in the inset of Fig. 15. Again, the full-wave results shown in Fig. 15

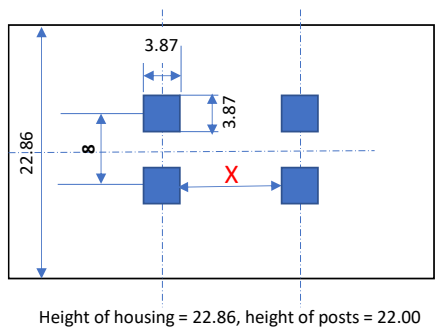

Height of housing = 22.86, height of posts = 22.00

Fig. 16. Configuration for investigation of inter-resonator couplings of dedicated dual-post resonances

confirm this conclusion.

Thus, the transvers dual-post constitutes a real singlet based on the odd mode resonance at the filter passband frequency and a bypass coupling of the evanescent $TE_{10}$ mode, similar to the waveguide implementation, which relies on $TE_{201}$ and $TE_{10}$ cavity modes [14]. Experimental validation of this general behavior has been provided by the examples in [5,6]. These configurations can be designed through standard filter design methods, considering the respective couplings of the dual-post resonator with the adjacent filter in/output ports and/or resonators; where the cross coupling value can be almost determined by the complete distance bypassing the dual-post. There is only a minor impact by the spurious basic mode. due to its large frequency distance. There is no advantage to a more elaborate higher order equivalent circuit model, which includes the spurious resonance that cannot be controlled.

*Inter-resonator coupling*

For the investigation of the inter-resonator couplings, the well-known resonance analysis is applied. First, a configuration with two identical dual-post units is explored by varying the separation distance between them (cf. Fig. 16).

Obviously, the resonance analyses yield four resonance frequencies; two of which are dedicated to the dominant dual-post mode and the other two to the 2nd mode. The deduced coupling values of the assigned modes are shown in Fig 17. The coupling between the dominant modes is substantially stronger (more than 5 times) that between the 2nd (odd) modes. The analyses of half the configuration in Fig. 16 considering an electric wall in the longitudinal symmetry plane (i.e., half waveguide width with center offset single posts) yield identical results as for the 2nd mode. This evidently demonstrates that the coupling of the 2nd mode in this configuration solely relies on the $TE_{20}$ waveguide mode and not on the fundamental $TE_{10}$ one. There will be a certain increase of the coupling values when considering the single post resonators centered inside this half width waveguide envelop (cf. Fig. 17).

Another coupling analysis was carried out for the altered original configuration, considering only one post (with center offset) of each pair. In this case the post height is modified (20.85mm) to have the same resonance frequency as the 2nd mode. Although this configuration exhibits weaker coupling values than those achieved for the dominant mode of the dual posts, in comparison with the 2nd mode couplings, they are more than three times stronger (cf. Fig. 17).

The method for this coupling analyses is based on identical resonating configurations. Consequently, the application to different resonator types is critical (as in the combination of a dual-post with a single one). However, the analyses of initial filter designs have shown that couplings between dual- and single post resonators are in the same order of magnitude as that obtained for the 2nd mode dual-post ones. This is attributed to the fact that the 2nd dual-post mode can be only coupled by odd mode field components, i.e., couplings with single post resonators in this configuration are only possible if these exhibit a center offset relative to the dual post.

An assessment of the footprint dimensions of the 3-order filter provided in [6, Fig. 6] confirms the above analyses. Indeed, the coupling of the single-post resonators with the dual-post (odd-mode) resonator with a close distance is dedicated to the $TE_{20}$ waveguide mode, while the bypass coupling between the two single posts can be (almost entirely) attributed to the $TE_{10}$ and results in much stronger coupling despite the longer distance.

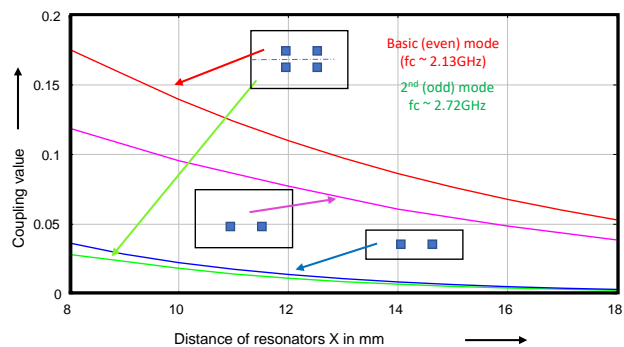

Fig. 17. Coupling values of inter-resonator couplings; dual-post configuration, basic mode red; 2nd mode green; magenta: single post configuration with same envelop configuration and center offset, post height (20.85mm) adjusted ($f_c$=2.72GHz); blue: centered single posts in half width envelop.

*Inline dual-post doublet*

The basic understanding of the physics of close post resonator configurations reveals many new design possibilities to be included in advanced filter designs. The knowledge that the 2nd mode of transverse dual-post resonators is always coupled by the $TE_{20}$ mode, while the $TE_{10}$ mode provides a strong bypass coupling has been considered by the singlet design above. The singlet configuration can also be extended by a second dual-post resonator for the realization of an inline second order building block, providing a pair of transmission zeros as shown in the example in Fig. 18. Also in this configuration, the opposite offsets of input and output couplings

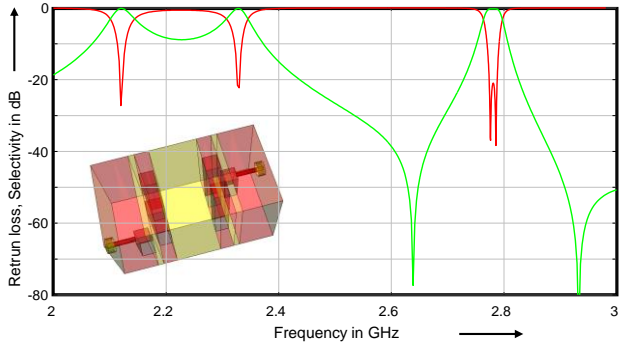

Fig. 18. Dual-post doublet; analyzed response with opposite offset spacing of coupling stubs (cf. inset)

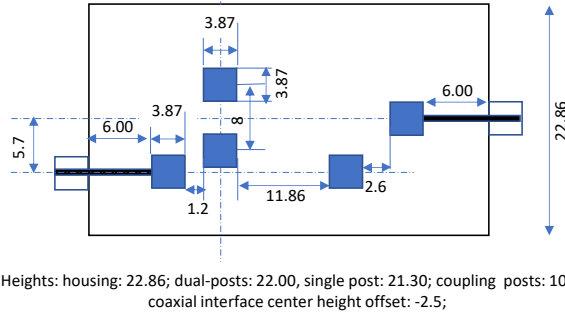

Heights: housing: 22.86; dual-posts: 22.00, single post: 21.30; coupling posts: 10.00; coaxial interface center height offset: -2.5;

Fig. 19. Initial filter configuration after assessment of resonator, input/output and inter-resonator coupling dimensions.

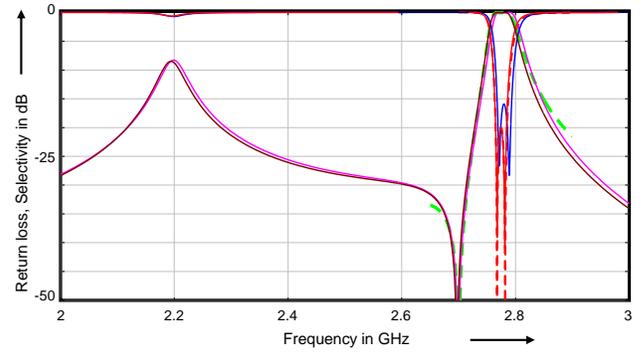

Fig. 20. 2-order filter design with dual- and single post resonators; synthesized characteristic: dashed lines; analyzed response of initial configuration in Fig. 19; blue: return loss, magenta: selectivity; results after minor tuning of coupling $M_{12}$ and resonators red: return loss, green: selectivity.

are important for the coupling sign transformation ($M_{SL}$). Obviously, the response exhibits two spurious resonances below the passband. The use of standard filter design methods allows the consideration of such basic configurations for advanced filter implementations. For example the inline dual-post doublet can be combined arbitrarily with other (e.g., single post) resonators to provide an inline quadruplet in higher order filter designs.

*2-order filter design example*

The following example introduces the systematic filter design that relies on well-established methods while considering an implementation with dual- and single post resonators. It allows also the identification of design limitations, e.g., resulting from the general $TE_{10}$ bypass couplings when using the 2$^{nd}$ dual-post mode, i.e., there is always a bypass coupling.

The 2-order filter characteristic shall have a center frequency of 2.775GHz, a bandwidth of 20MHz (return loss: 20dB) and a TZ at 2.7GHz. This response is satisfied by the normalized coupling matrix:

$$M = \begin{bmatrix} 0 & 1.2074 & -0.2443 & 0 \\ 1.2074 & 0.3559 & 1.6116 & 0 \\ -0.2443 & 1.6116 & -0.2973 & 1.2300 \\ 0 & 0 & 1.2300 & 0 \end{bmatrix}$$

The filter implementation considers the 2$^{nd}$ mode of a dual post and a single post resonator. Thus, the first design step is related to the determination of the two different resonator types as described above. The second step is related to the design of the input and output coupling of the assigned resonators. As shown above, the maximum coupling of the dual-post resonator is achieved at a transverse offset of 5.7mm. Hence, the required coupling is adjusted by the longitudinal distance between coupling stub and resonator. The configuration of the single post output coupling considers a centered location of the coupling stub, while the post is transversally offset by 5.7mm. (The post offset accounts for the later inter-resonator coupling with the 2$^{nd}$ mode dual-post resonator, while the central coupling location avoids a parasitic coupling with the 2$^{nd}$ dual post resonance mode). The adjustment of the required output coupling is made in the same manner, i.e., by varying the longitudinal distance of the stub.

Once the configurations of the required interface couplings have been determined, they can be combined for the 2-order filter with the intermediate coupling length. Based on the considerations above, the initial length (11.86mm) for the required coupling is estimated by two single post resonators in the half width waveguide envelope. The filter dimensions obtained by these initial systematic steps are given in Fig. 19.

The response (cf. Fig. 20) of this initial configuration exhibits a 2-order filter with the required bandwidth, 17dB return loss, and a frequency offset of about 4MHz. Thus, this pre-dimensioning which relies on well-established filter design methods need evidently only minor adjustments of resonators and couplings to meet the original specifications as shown by the analyzed results of the final configuration in Fig. 20.

Note, apart from the related offset locations of the input coupling and the single-post resonator, the cross coupling has not been considered for this initial design. Due to the above evaluation there is always a bypass coupling of the 2$^{nd}$ mode of the dual-post resonator in such structures by the dominant $TE_{10}$ waveguide mode, since this mode does not interact with the dual post resonator but with interface couplings and single post resonators. For the present configuration, the bypass coupling results mainly from the distance of the input coupling to the single post, while the couplings of dual post resonator (towards input and single post) also depend on the associated sub-lengths. A TZ location closer to the passband requires a stronger bypass coupling while the other couplings have to be remain with a similar strength. Although all dedicated couplings depend on the respective longitudinal distances, they can be adjusted individually according to the need for a desired filter function (TZ location). For example, the bypass coupling $M_{s2}$ in the example is increased by reducing the distance between the dual- and single post resonators; to keep the respective main coupling $M_{12}$ (almost) constant, the transverse center offset of the single post resonator location has to be reduced (see investigations above). Consequently, the knowledge of the real functionality supports the systematic straight forward filter designs using well-establish methods.

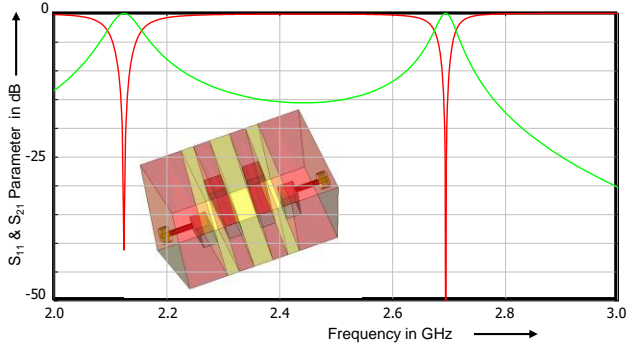

Fig. 21. Analyzed response of in-line dual-post resonator; (configuration, cf. inset)

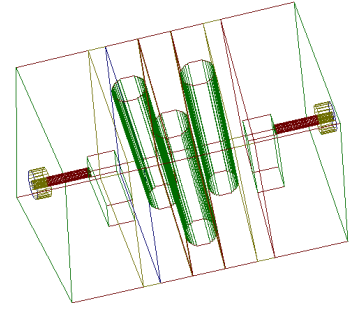

Fig. 22. Triple-post (doublet) filter configuration using coupling stubs at both sides

## VII. ADDITIONAL VERIFICATION EXAMPLES

### A. In-line dual-post resonator

First, an in-line dual-post unit is considered. The input and output are aligned with the posts of the unit as shown in the inset of Fig. 21. As discussed earlier, this unit is not able to generate a TZ at finite frequency. This is confirmed by the analyzed full-wave results shown in Fig. 21. This behavior is identical to that of $TE_{101}$ and $TE_{102}$ waveguide cavity modes in in-line filters (cf. [4]).

### B. Triple-post unit (doublet)

As discussed earlier, the triple-post unit is capable of generating up to two TZ's at finite frequencies. Only one TZ is generated when the coupling coefficients are constrained according to equation (10.c). Even if this constraint is not satisfied exactly, one TZ will be located far away from the passband, especially when the dominant even mode resonates at a sufficiently distant frequency below the passband. It was also demonstrated that the remaining TZ satisfies an approximate TZ-shifting property.

The triple-post configuration in Fig. 6 is used with necessary input and output coupling, symmetrically placed at both sides, to realize a 2$^{nd}$ order (doublet) filter. (Note, the footprint and housing cross sections are according to Fig.12, [7]). For the input and output of the filter we use coupling stubs with a conductive connection to the interfacing coaxial lines (cf., Fig. 22). The stub location and height are optimized to achieve a 2$^{nd}$ order filter characteristic using the triple-post dimensions of Fig. 6.

The obtained characteristic is shown in Fig. 23. The configuration provides an equiripple passband from 2.747GHz to 2.759GHz (return loss 22dB) with a transmission zero above the passband at 2.767GHz. By *solely* adjusting the heights of the posts we get an almost mirrored image of the response with the transmission zero moved below the passband. The spurious mode far below remains practically unchanged (see Figs. 23,24). In detail, the height of the in-line dual posts of the triple configuration is reduced to 21.912mm, whereas the height of single post is increased to 22.042mm. Fig. 25 depicts both responses over a frequency range around the passband. This clearly demonstrate the principle doublet behavior with the TZ-shifting property. Although the possibility of shifting the TZ by adjusting the resonant frequencies was mentioned in [7], the equivalent circuit used in the design does not exhibit this property.

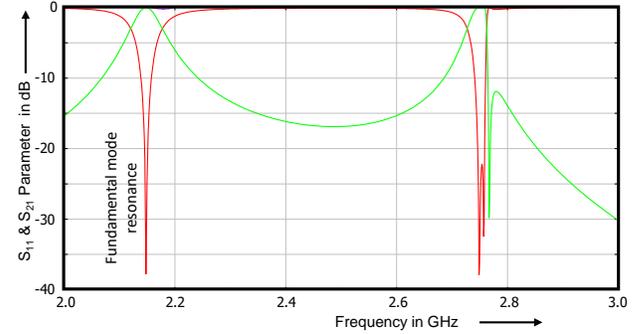

Fig. 23. Response of configuration in Fig. 14 (height all posts 21.95mm)

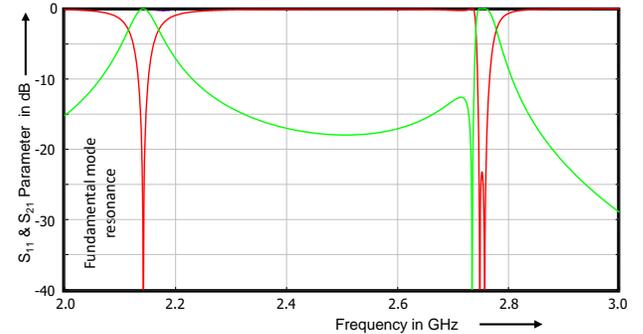

Fig. 24. Response of configuration in Fig. 14 (post heights: inline pair 21.912mm, single post 22.042mm)

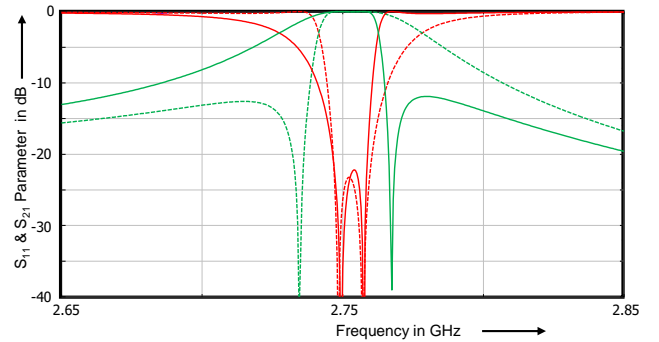

Fig. 25. Narrow-band responses of initial doublet filter (Fig. 23,24) and with solely adjusted resonator post heights (zero-shifting property)

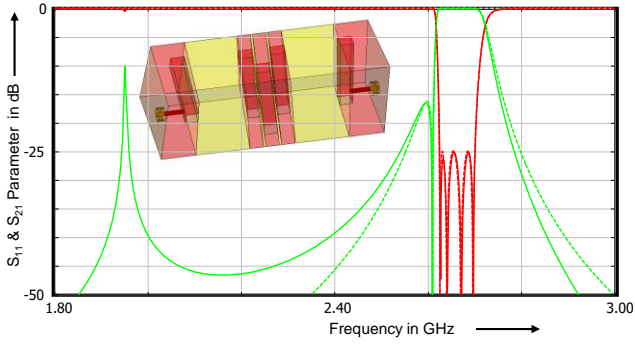

Fig. 26. 4-order filter design; dashed lines – synthesis of 'box-section' characteristic; solid lines – analyzed responses of implementation (shown in inset).

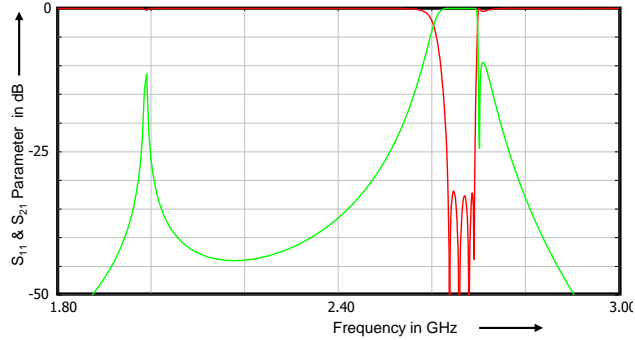

Fig. 27. 4-order filter design according inset Fig. 18; TZ shifting above passband by solely small adjustment of post heights

These results show that triple-post structures are adequately described by models that rely only on those resonances that contribute to the passband. Their design can be carried out by following well-established filter synthesis and design methods. There is no real need for elaborate overdetermined models and synthesis techniques that include spurious resonances that occur far away from the passband, cannot be controlled, and contribute very little to the passband.

### C. Design of a 4$^{th}$ order box-section filter

A 4-order filter implementation considers a triple-post doublet section as basic building block, coupled at both sides with single post resonators to realize a 4-order filter with one TZ. The filter response has a center frequency at 2.662GHz the bandwidth is 76MHz (equiripple return loss 25dB) and a TZ below the passband at 2.61GHz. According to the above findings the configuration corresponds to a box-section coupling scheme, i.e., the respective normalized coupling values for the filter response are given by[7]:

$$M = \begin{pmatrix} 0 & 1.1402 & 0 & 0 & 0 & 0 \\ 1.1402 & 0.0964 & 0.9398 & 0.4229 & 0 & 0 \\ 0 & 0.9398 & 0.3934 & 0 & -0.9398 & 0 \\ 0 & 0.4229 & 0 & -1.0211 & 0.4229 & 0 \\ 0 & 0 & -0.9398 & 0.4229 & 0.0964 & 1.1402 \\ 0 & 0 & 0 & 0 & 1.1402 & 0 \end{pmatrix} \quad (13)$$

The dashed lines in Fig. 26 depict the response of the synthesized coupling matrix. The implementation is shown by the inset in Fig.26; the central triple-post doublet and the single resonators are situated in a straight housing (similar to the case in [7]). The single posts are conductively couplet with the coaxial interfaces at the opposite end walls. The different couplings between the single posts and the triple-post resonances are adjusted by the transverse offset locations of the single-posts and the inline-distance. The resonances of the doublet are conveniently controlled by separate adjustment of the dedicated two inline posts and the single post beside the two. The analyzed characteristics of the configuration are shown in Fig. 26; they coincide accurately with the ideal response of the coupling matrix. As expected, the spurious fundamental mode resonance of the triple-post unit appears far below the passband resulting in an impairment of the ideal ('box-section') filter response with increasing distance from the passband. Towards lower frequencies there is a reduced rejection, while towards higher frequencies an increase of rejection is observed.

It should be noted that this design also exhibits the principle 'zero-shifting' property as demonstrated by the response in Fig. 27, which follows from Fig. 26 by solely adjusting the heights of the posts. As the analyzed results show, the fundamental mode resonance still remains far below the passband. However, the achieved response does not provide an accurate mirror image of the initial design (Fig.26); i.e., the passband is narrower with a higher return loss (>30dB) and the TZ is closer to the band. This is attributed to the spurious effects described above and requires some minor adjustments of the couplings to achieve the accurate mirror image of the initial response (Fig. 26).

### C. Triplet Filter Design with In-Line Dual Post

An inline dual-post resonator (Fig. 2, left) was selected for a triplet filter design to further validate its basic properties. The dual-post configuration is coupled to single post resonators in a triangular arrangement with the latter resonators connected via conductive couplings with coaxial interfaces (cf. Fig. 28). For the cross coupling, an inductive window is considered in the separation wall between the single post resonators. The passband of the 3$^{rd}$ order filter characteristic is from 2.598GHz to 2.680GHz (equiripple return loss: 17dB) and a TZ is located below the passband at 2.555GHz. The dashed lines in Fig. 28 show the response of the synthesized coupling matrix (inset in Fig. 28) for this implementation. Note, for the realization of the TZ below the passband a transformation of the coupling sign is mandatory and therefore the utilization of the 2$^{nd}$ (odd) mode.

---

[7] The synthesis represents the narrow band characteristic close to the passband without any spurious effects.

The dual-post height in the final design is 22.06mm. The analyzed response of this configuration is shown in Fig. 29 (solid lines), together with the ideal response. Good agreement between the two is achieved. The analysis over a wide frequency range exhibits the spurious fundamental resonance far below the passband as expected (cf. Fig. 30).

The particular properties of the special dual-post resonator configuration (Fig. 28) are simply demonstrated by exchanging the 2$^{nd}$ (odd) mode and the fundamental mode allocations. That is, the resonances of both dual-post modes have been moved towards higher frequencies by reducing solely the height of the dual-posts in the structure (Fig. 28). The analysis of this identical configuration with only a reduced dual-post height of 20.63mm exhibits a transformation of the coupling sign. The computed response (Fig. 31) exhibits a 3$^{rd}$-order filter response with moderate return loss (~8dB), but with the TZ above the passband; additionally, the odd mode is now the spurious one far above the passband (cf., Fig. 31). No attempt was made to improve the return loss of a typical value of 20 dB say because the point here is to show the sign transformation that comes with only adjusting the heights of the posts of the dual-post unit. The simple interchange of even and odd modes for a dedicated filter implementation will not yield a mirror image of the responses. The modes exhibit different field strengths for the couplings with the other resonators (posts). Note that this feature is different from the zero-shifting property of doublet configurations that involve the frequency interchange of the assigned resonances ('self-couplings, $M_{xx}$'). Instead, this involves the interchange of the roles played by the two resonances of the dual-post unit. In the first design, the even fundamental mode is spurious and located below the passband. In the second design, the odd mode is spurious and located above the passband. This operation allows different transformations of the coupling coefficients and extends the design possibilities.

The additional TZ that is present in these two filters can be explained by the interference of the signal going through the coupling iris between the posts at the input and the output and the spurious resonance. In the first case (Fig. 30), the bypassed mode is even. Around the spurious resonance, the filter acts as a singlet with all coupling coefficients of the same sign, hence the TZ above the resonance. In the second case (Fig. 31), the spurious mode is odd. This means that one coupling coefficient is negative. Around the spurious resonance, the filter is equivalent to a singlet with one negative coupling coefficient, hence the TZ is below the resonance.

## VIII. REFERRED EXPERIMENTAL VALIDATION

Measured results of filter designs with close post configurations have been provided in several recent papers [5-9]. Photos of hardware realizations were also provided in the same references. The main goal of this paper is not to design and manufacture new filters based on the same structures; instead, it is to explain the underlying physics of the relevant basic building blocks to allow straight-forward systematic design and tuning by using well-established filter design methods. Moreover, it will be the key for many novel and innovative design ideas.

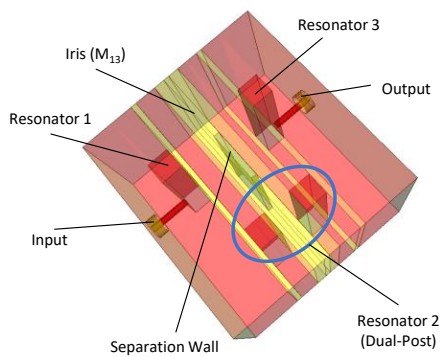

Fig. 28. Configuration of triplet filter with dual-post resonator

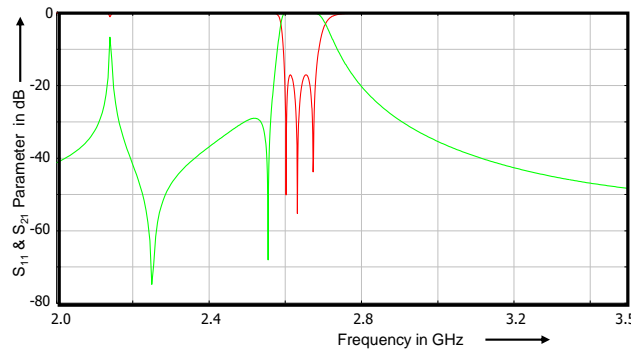

Fig. 30. Analyzed wideband response of triplet filter design Fig. 27.

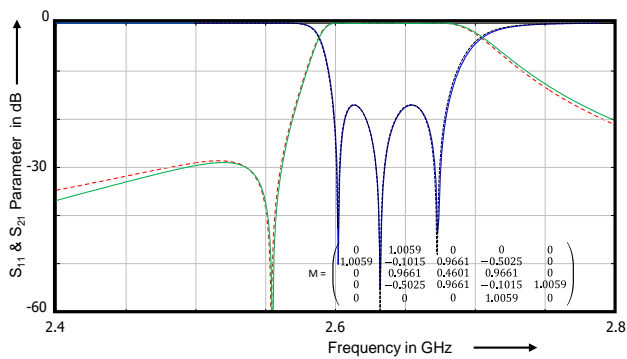

Fig. 29. Responses of triplet filter design (Fig. 28); synthesis (dashed), analyses (solid); inset: coupling matrix

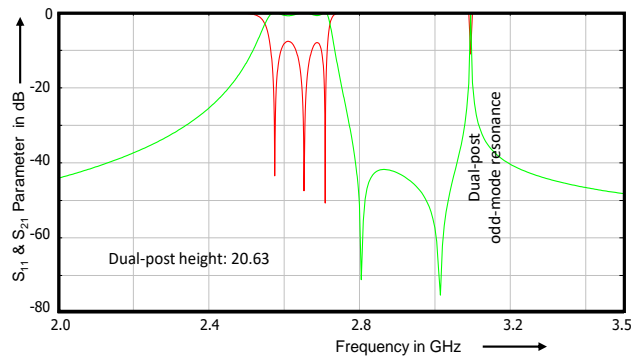

Fig. 31. Analyzed wide response of triplet filter design in Fig.27 but reduced height of dual-post unit.

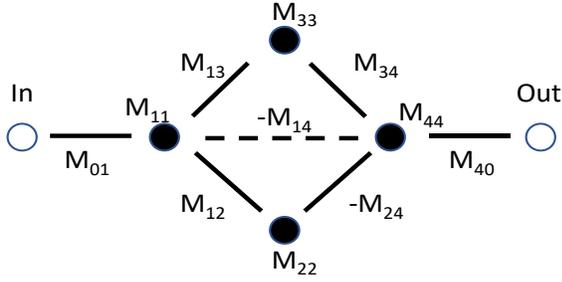

Fig. 32. Real coupling scheme of introduced 4-pole filter ([7] Figs. 12-15) representing physical resonances of configuration ($M_{34}=M_{13}$; $M_{24}=M_{12}$; $M_{44}=M_{11}$)

Obviously the elaborate and questionable syntheses in combination with EM-optimization techniques in [5-9] produce working filters. The point we make in this paper is that the same filters can be designed straightforward by using classic design methods that are familiar to all filter designers if the proper equivalent circuits are used. We use the example of a 4-th order filter to illustrate this point.

The 4-order filter design in [7] is similar to the 4$^{th}$ order filter design introduced in section VII, C. Because of the orthogonality of the solutions of Maxwell's equations in a given volume, the eigen-resonances of the triple-post unit are not coupled to each other. Given that the bandwidth of the filter is not large, the coupling coefficients of the resonances of the triple-post unit to the input and output resonators are not strong. This means that we can treat them as localized resonators. The spurious (fundamental) mode provides a coupling between the input and output resonators that is small and practically constant over the narrow passband. The result is the coupling scheme (box-section with small bypass coupling) shown in Fig. 32. A comparison of the filter response shown in ([7], Fig. 15) with a synthesized characteristic considering one TZ only (box-section without bypass), exhibits some small deviations of the rejection; i.e., the rejection above the passband is few dB higher and below few dB lower. This is directly attributed to the weak spurious fundamental mode bypass coupling in the box-section scheme (Fig. 32), that yields an extra TZ far above the passband; i.e., the corresponding normalized coupling values are given by equation (14).

$$M = \begin{pmatrix} 0 & 1.0390 & 0 & 0 & 0 & 0 \\ 1.0390 & 0.1029 & 0.8984 & 0.2365 & -0.1046 & 0 \\ 0 & 0.8984 & 0.1143 & 0 & -0.8984 & 0 \\ 0 & 0.2365 & 0 & -1.0136 & 0.2365 & 0 \\ 0 & -0.1046 & -0.8984 & 0.2365 & 0.1029 & 1.0390 \\ 0 & 0 & 0 & 0 & 1.0390 & 0 \end{pmatrix} \quad (14)$$

Note that all the elements of this coupling matrix have typical values contrary to the unrealistic values that result when using the triplet resonator scheme representations in [7,9].

It should be noted, that the weak spurious mode coupling is an inherent parasitic effect in this doublet configuration with the triple-post structure; its accurate strength depend on the final physical implementation (mainly on the distance of the spurious fundamental mode resonance from the passband). It cannot be controlled completely as part of the design. It makes little engineering sense to include it in the model upon, which the design is based. The best that could be done is to push it as far away from the passband as possible by reducing the distance between the strongly interacting posts.[8] However, the knowledge of its effect from a first iteration may be incorporated to a certain extent in the initial synthesis of such filter design. Final adjustments of the relevant dimensions of the filter complete the design to compensate for the small distortions that may be caused by the spurious resonance.

Final note: The above investigations (*section VII, B*) regarding the properties of the triple-post resonator implementation are identical with the central footprint and housing dimensions in [7] Fig. 12, that are part of the realized 4-order filter design in [7] (photo of filter Fig. 14, measured results Fig. 15). However, the passband of the realized filter in [7] is substantially lower than the resonance frequencies of the presented basic investigations. This is of course attributed to the different height and loading of the posts (height of posts in [7] is given as 22.86mm with additional dielectric loading at the top, without further information). Nevertheless, this will not change the principle physical properties of such close triangular post configurations.

Also the experimental results provided in [9] validate the real physical behavior of close post configurations in coaxial filter designs introduced above, i.e., all these filter designs rely on standard synthesis methods where the number of resonators within the configuration is identical with that of the filter order. Consequently, also the introduced synthesis in [9] is misleading, it introduces extra 'redundant resonances', but without giving any physical representation of such redundant resonance circuits, but a filter characteristic will be finally achieved using commercial EM-CAD optimization techniques.

## IX. CONCLUSIONS

The investigations of building blocks with strongly coupled posts have shown that the individual posts are not pivotal to the design of filters containing these structures. Because of the strong coupling between posts that share the same volume, resonances localized around the post are not physical. Instead, the resonances of the complete block and their properties have to be examined to identify those features that are relevant to the design of filters that contain one or more of such structures. Similarity transformations yield coupling matrices that conserve the frequency response of the entire filter but not necessarily the dominant 'internal' physics of these structures because of the strong coupling and the boundary conditions. Nevertheless, coupling matrices based on localized resonance can be and have been successfully used to optimize the filter in connection with its *full-wave* response of the entire filter or subsections, that consider these structures with adjacent single post resonators. However, classic design methods based on pairs of resonators will most likely fail when applied to such coupling matrices. Using resonances that satisfy the boundary conditions and contribute to the passband is essential to understanding how these structures work and how to use them in interesting and innovative designs. These filters can be designed by following well-established methods as long as they are based on an equivalent circuit (coupling matrix) that contains only those physical resonances that contribute to the passband.

---

[8] The basic knowledge of the functionality of this triple-post dual-mode resonator allows similar basic design consideration as provided example for the dual-post singlet resonator introduced in section VI, e.g., to optimize unloaded Q and spurious mode distance, as well as coupling properties with other resonator types.

The knowledge of the real functionality is the prerequisite for the systematic application of the standard design methods for a convenient geometrical pre-design of the desired filter configuration, considering also different kinds of resonators without general design limitations. As an example, the usual basic considerations for the pre-design of a 2-order filter with one TZ are introduced for an implementation using dual- and single post resonators. The analysis results of the initial configuration, obtained by the application of this classical design approach, provides almost the desired filter characteristic response without any EM-optimization/tuning.

The filters presented in this paper have all been designed by the well-established method. The effect of spurious resonance is then taken into account by a final adjustment of the dimensions of the filter in connection with the equivalent circuit and a field solver.